\documentclass[%
reprint,
superscriptaddress,
 amsmath,amssymb,
aps,
prb,
]{revtex4-2}

\usepackage{graphicx}
\usepackage{dcolumn}
\usepackage{bm}
\usepackage{hyperref}
\usepackage[capitalise]{cleveref}
\hypersetup{
    colorlinks = true,
    allcolors=black,
    citecolor=blue,
    linkcolor=blue
} 
\usepackage{amsmath, amsfonts, dsfont, mathrsfs, float,amssymb}
\usepackage{color}
\usepackage[caption=false]{subfig}
\usepackage{svg}
\usepackage{tikz}
\usepackage{tikz-network}
\usepackage{orcidlink}
\usepackage{xcolor}
\usepackage{mathtools}
\usepackage{multirow}
\usepackage{csquotes}
\usepackage{svg}

\begin{document}


\title{Can Quantum Computers Do Nothing?} 

\author{Alexander Nico-Katz}
\email{nicokata@tcd.ie}
\affiliation{School of Physics, Trinity College Dublin, Dublin 2, Ireland}
\affiliation{Trinity Quantum Alliance, Unit 16, Trinity Technology and Enterprise Centre, Pearse Street, D02 YN67, Dublin 2, Ireland}
\author{Nathan Keenan}
\email{nakeenan@tcd.ie}
\affiliation{School of Physics, Trinity College Dublin, Dublin 2, Ireland}
\affiliation{Trinity Quantum Alliance, Unit 16, Trinity Technology and Enterprise Centre, Pearse Street, D02 YN67, Dublin 2, Ireland}
\affiliation{IBM Quantum, IBM Research Europe - Dublin, IBM Technology Campus, Dublin 15, Ireland}
\author{John Goold}
\email{gooldj@tcd.ie}
\affiliation{School of Physics, Trinity College Dublin, Dublin 2, Ireland}
\affiliation{Trinity Quantum Alliance, Unit 16, Trinity Technology and Enterprise Centre, Pearse Street, D02 YN67, Dublin 2, Ireland}
\affiliation{Algorithmiq Ltd, Kanavakatu 3C 00160, Helsinki, Finland}

\date{\today}
             
\begin{abstract}
    Quantum computing platforms are subject to contradictory engineering requirements: qubits must be protected from mutual interactions when idling (`doing nothing'), and strongly interacting when in operation. If idling qubits are not sufficiently protected, information can `leak' into neighbouring qubits, become non-locally distributed, and ultimately inaccessible. Candidate solutions to this dilemma include patterning-enhanced many-body localization, dynamical decoupling, and active error correction. However, no information-theoretic protocol exists to actually quantify this information loss due to internal dynamics in a similar way to e.g. SPAM errors or dephasing times. In this work, we develop a scalable, flexible, device non-specific protocol for quantifying this bitwise idle information loss based on the exploitation of tools from quantum information theory. We implement this protocol in over 3500 experiments carried out across 4 months (Dec 2023 - Mar 2024) on IBM's entire Falcon 5.11 series of processors. After accounting for other sources of error, and extrapolating results via a scaling analysis in shot count to zero shot noise, we detect idle information leakage to a high degree of statistical significance.  This work thus provides a firm quantitative foundation from which the protection-operation dilemma can be investigated and ultimately resolved.
\end{abstract}

\maketitle

\section{Introduction}
\label{sec:intro}

\begin{figure}[!ht]
    \centering
    \begin{tikzpicture}[scale=0.85]

        \Text[x=-1.2, y=5.5, fontsize=\Large]{\textbf{(a)}}
        \Text[x=-1.2, y=3.05, fontsize=\Large]{\textbf{(b)}}
        \Text[x=-1.2, y=-0.7, fontsize=\Large]{\textbf{(c)}}
    
        \Vertex[x=0.0, y=5.5, Pseudo]{idealcircA}
        \Vertex[x=7.0, y=5.5, Pseudo]{idealcircB}
        \Vertex[x=1.0, y=5.8, Pseudo]{idealcircL0}
        \Vertex[x=0.8, y=5.2, Pseudo]{idealcircL1}
        \Edge(idealcircA.center)(idealcircB.center)
        \Edge(idealcircL0.center)(idealcircL1.center)

        \Vertex[x=0.0, y=2.0, Pseudo]{circA0}
        \Vertex[x=7.0, y=2.0, Pseudo]{circA1}
        \Vertex[x=0.0, y=2.7, Pseudo]{circB0}
        \Vertex[x=7.0, y=2.7, Pseudo]{circB1}
        \Vertex[x=0.0, y=3.4, Pseudo]{circC0}
        \Vertex[x=7.0, y=3.4, Pseudo]{circC1}
        \Vertex[x=0.0, y=4.1, Pseudo]{circD0}
        \Vertex[x=7.0, y=4.1, Pseudo]{circD1}

        \Vertex[x=2.3333, y=2.0, Pseudo]{circAM0}
        \Vertex[x=4.6666, y=2.0, Pseudo]{circAM1}

        \Vertex[x=3.5, y=2.7, Pseudo]{circBM0}

        \Vertex[x=1.75, y=4.1, Pseudo]{circDM0}
        \Vertex[x=3.5, y=4.1, Pseudo]{circDM1}
        \Vertex[x=5.25, y=4.1, Pseudo]{circDM2}
        
        \Edge[bend=10](circA0.center)(circAM0.center)
        \Edge[bend=-10](circAM0.center)(circAM1.center)
        \Edge[bend=10](circAM1.center)(circA1.center)

        \Edge[bend=-6](circB0.center)(circBM0.center)
        \Edge[bend=6](circBM0.center)(circB1.center)
        
        \Edge[bend=-5](circC0.center)(circC1.center)

        \Edge[bend=15](circD0.center)(circDM0.center)
        \Edge[bend=-15](circDM0.center)(circDM1.center)
        \Edge[bend=15](circDM1.center)(circDM2.center)
        \Edge[bend=-15](circDM2.center)(circD1.center)

        \Vertex[x=2.0, y=3.3, Pseudo]{vertI0}
        \Vertex[x=2.0, y=4.1, Pseudo]{vertI1}
        \Edge[style={dashed}](vertI0.center)(vertI1.center)

        \Vertex[x=5.0, y=2.79, Pseudo]{vertI0}
        \Vertex[x=5.0, y=4.1, Pseudo]{vertI1}
        \Edge[style={dashed}](vertI0.center)(vertI1.center)

        \Vertex[x=1.3, y=3.3, Pseudo]{vertI0}
        \Vertex[x=1.3, y=2.622, Pseudo]{vertI1}
        \Edge[style={dashed}](vertI0.center)(vertI1.center)

        \Vertex[x=4, y=2.72, Pseudo]{vertI0}
        \Vertex[x=4, y=1.9, Pseudo]{vertI1}
        \Edge[style={dashed}](vertI0.center)(vertI1.center)

        \Vertex[x=3.5, y=4.1, Pseudo]{vertI0}
        \Vertex[x=3.5, y=3.2, Pseudo]{vertI1}
        \Edge[Direct, bend=20](vertI0.center)(vertI1.center)
        \Edge[Direct, bend=20](vertI1.center)(vertI0.center)

        \Vertex[x=0.8, y=3.35, Pseudo]{vertI0}
        \Vertex[x=0.8, y=2.08, Pseudo]{vertI1}
        \Edge[Direct, bend=20](vertI0.center)(vertI1.center)
        \Edge[Direct, bend=20](vertI1.center)(vertI0.center)

        \Vertex[x=3.0, y=3.2, Pseudo]{vertI0}
        \Vertex[x=3.0, y=1.9, Pseudo]{vertI1}
        \Edge[Direct, bend=20](vertI0.center)(vertI1.center)
        \Edge[Direct, bend=20](vertI1.center)(vertI0.center)

        \Vertex[x=6.6, y=3.35, Pseudo]{vertI0}
        \Vertex[x=6.6, y=2.72, Pseudo]{vertI1}
        \Edge[Direct, bend=20](vertI0.center)(vertI1.center)
        \Edge[Direct, bend=20](vertI1.center)(vertI0.center)

        \Vertex[x=5.9, y=2.8, Pseudo]{vertI0}
        \Vertex[x=5.9, y=2.1, Pseudo]{vertI1}
        \Edge[Direct, bend=20](vertI0.center)(vertI1.center)
        \Edge[Direct, bend=20](vertI1.center)(vertI0.center)

        \fill [blue, opacity=0.4] (5.25,-1.75) rectangle (7.35,-1.05);
        \fill [blue, opacity=0.4] (5.95,-1.05) rectangle (6.65,-0.35);

        \fill [orange, opacity=0.5] (0.35,0.35) rectangle (1.05,-0.35);
        \fill [orange, opacity=0.5] (1.05,-1.05) rectangle (1.75,-1.75);
        \fill [orange, opacity=0.5] (4.55,-1.75) rectangle (5.25,-2.45);
        \fill [orange, opacity=0.5] (5.95,-0.35) rectangle (6.65,0.35);

        \Vertex[x=0.0, y=0.0, color=blue, size=.4]{q0}
        \Vertex[x=0.7, y=0.0, color=blue, size=.4]{q1}
        \Vertex[x=1.4, y=0.0, color=blue, size=.4]{q4}
        \Vertex[x=2.1, y=0.0, color=blue, size=.4]{q7}
        \Vertex[x=2.8, y=0.0, color=blue, size=.4]{q10}
        \Vertex[x=3.5, y=0.0, color=blue, size=.4]{q12}
        \Vertex[x=4.2, y=0.0, color=blue, size=.4]{q15}
        \Vertex[x=4.9, y=0.0, color=blue, size=.4]{q18}
        \Vertex[x=5.6, y=0.0, color=blue, size=.4]{q21}
        \Vertex[x=6.3, y=0.0, color=blue, size=.4]{q23}

        \Vertex[x=0.7, y=-1.4, color=blue, size=.4]{q3}
        \Vertex[x=1.4, y=-1.4, color=blue, size=.4]{q5}
        \Vertex[x=2.1, y=-1.4, color=blue, size=.4]{q8}
        \Vertex[x=2.8, y=-1.4, color=blue, size=.4]{q11}
        \Vertex[x=3.5, y=-1.4, color=blue, size=.4]{q14}
        \Vertex[x=4.2, y=-1.4, color=blue, size=.4]{q16}
        \Vertex[x=4.9, y=-1.4, color=blue, size=.4]{q19}
        \Vertex[x=5.6, y=-1.4, color=blue, size=.4]{q22}
        \Vertex[x=6.3, y=-1.4, color=blue, size=.4]{q25}
        \Vertex[x=7.0, y=-1.4, color=blue, size=.4]{q26}

        \Vertex[x=0.7, y=-0.7, color=blue, size=.4]{q2}
        \Vertex[x=3.5, y=-0.7, color=blue, size=.4]{q13}
        \Vertex[x=6.3, y=-0.7, color=blue, size=.4]{q24}

        \Vertex[x=2.1, y=0.7, color=blue, size=.4]{q6}
        \Vertex[x=4.9, y=0.7, color=blue, size=.4]{q17}
        \Vertex[x=2.1, y=-2.1, color=blue, size=.4]{q9}
        \Vertex[x=4.9, y=-2.1, color=blue, size=.4]{q20}

        \Vertex[x=0.7, y=0.0, shape = diamond, color=yellow, size=0.3]{targetRAND}
        \Vertex[x=6.3, y=-1.4, shape = diamond, color=yellow, size=0.3]{targetPLAQ}

        \Edge(q0)(q1)
        \Edge(q1)(q4)
        \Edge(q4)(q7)
        \Edge(q7)(q10)
        \Edge(q10)(q12)
        \Edge(q12)(q15)
        \Edge(q15)(q18)
        \Edge(q18)(q21)
        \Edge(q21)(q23)
        \Edge(q3)(q5)
        \Edge(q5)(q8)
        \Edge(q8)(q11)
        \Edge(q11)(q14)
        \Edge(q14)(q16)
        \Edge(q16)(q19)
        \Edge(q19)(q22)
        \Edge(q22)(q25)
        \Edge(q25)(q26)
        \Edge(q6)(q7)
        \Edge(q17)(q18)
        \Edge(q1)(q2)
        \Edge(q2)(q3)
        \Edge(q23)(q24)
        \Edge(q24)(q25)
        \Edge(q8)(q9)
        \Edge(q19)(q20)
        \Edge(q12)(q13)
        \Edge(q13)(q14)

    \end{tikzpicture}
    \caption{\textbf{Schematic quantum circuits and devices} \textbf{(a)}  Schematic of an ideal quantum circuit which realizes perfect idling. \textbf{(b)} Schematic of a physical quantum circuit with imperfect idling due to native transport (black arrows), interactions (dotted lines), and precession (oscillating lines). \textbf{(c)} Schematic showing the qubit geometry of the Falcon 5.11 series of IBM's quantum devices. Example target qubits are shown by yellow diamonds. Shaded regions respectively indicate examples of a contiguous set of nearest-neighbour qubits (blue) and a random set of qubits excluding nearest-neighbours (orange). Measuring the additional information gained about the target qubit when addressing the complementary qubits allows us to quantify idle information loss due to the internal dynamics of the hardware.}
    \label{fig:schematic}
\end{figure}
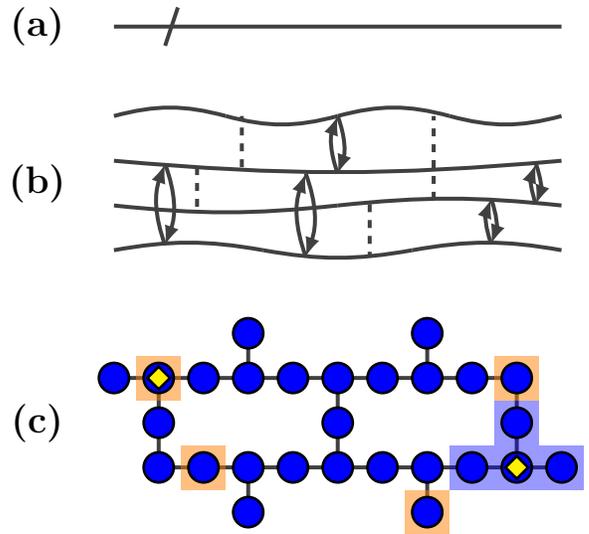

Quantum computing is a new paradigm of computation based on the exploitation of quantum phenomena which promises frontier impacts on global energy, health, materials science, and technological innovation \cite{altman2021quantum, bravyi2022future}. With the firm advent of the noisy intermediate-scale quantum (NISQ) era in the form of accessible quantum processors containing some tens to hundreds of qubits \cite{preskill2018quantum, arute2019quantum, chow2021ibm, kim2023evidence}, out-of-the-box error mitigation \cite{van2023probabilistic}, and the nascent implementation of simple error correction \cite{takeda2022quantum, google2023suppressing, xu2024constant}, this new paradigm is now being mapped out comprehensively. However, far from their conceptualization as ideal systems, real quantum computers are intrinsically programmable many-body quantum systems with complicated internal dynamics exposed to stray interactions and thermal fluctuations  (\cref{fig:schematic}\textbf{(a-b)} schematically show idealized and realistic `empty' circuits respectively) \cite{xu2024lattice}. The precise control of these complex many-body physical systems is ultimately the fundamental problem of quantum computing. 

This problem is encapsulated by the `protection-operation dilemma' wherein qubits need to be coupled and decoupled from each other at different times during the same computation \cite{silveri2022many}. When not in active operation (`doing nothing'), idling qubits should be well protected from mutual interactions with other qubits in the device. However, multi-qubit operation require strong mutual interactions such that entanglement and correlations can be generated between the active qubits quickly. Both of these processes should also be resistant to environmental decoherence, which typically reduces to the requirement that computations should be completed as quickly as possible. This generates contradicting engineering requirements: qubits should be decoupled when idling, but strongly coupled when necessary for computation. The two main ways this contradiction is resolved are (i) many-body localizing the qubits through spatial disorder and only bringing neighbouring qubits into resonance during gate operations (ii) rapidly tuning the mutual interactions in-situ to actively couple and decouple qubits during runtime \cite{silveri2022many, qian2023mitigating, varvelis2024perturbative}. The potential breakdown of the former resolution has been addressed in detail by recent works which find that current-generation superconducting quantum computers may enter chaotic regimes during operation; calling into question how well protected qubits in these devices actually are \cite{berke2022transmon, borner2023classical, basilewitsch2024chaotic}. The latter poses a complicated engineering problem that may induce higher order or time-dependent effects which are not yet fully understood. Idle information leakage can also be reduced by active error correction or dynamical decoupling, but these introduce formidable engineering problems and additional gate-based errors and complexity respectively \cite{devitt2013quantum, ezzell2023dynamical}.

Inspired by the burgeoning many-body perspective on quantum computation, we exploit tools from many-body physics and quantum information theory to address a critical question underlying the entire protection-operation dilemma: can the information loss induced by multi-qubit dynamics during idling be quantified in a similar way to e.g. readout errors or dephasing times? In this work we address this question directly by presenting and experimentally implementing a protocol for quantifying idle information loss. This protocol, developed and discussed in \cref{sec:quantifying}, leverages the Holevo quantity to accurately quantify information loss, is device non-specific, scalable, and can be easily run during computational downtime. We experimentally implement the protocol in \cref{sec:experimental} across the entire range of Falcon 5.11 series of IBM's quantum devices \footnote{IBM has recently discontinued this particular series, but the protocol is not device-specific.}. After accounting for other sources of error, we identify a small but measurable amount of information leakage to a high degree of statistical significance. Our accurate quantification of the actual informational impact of many-body effects in real devices represents a decisive step towards identifying and measuring idle information loss in future quantum devices. This result also provides a firm quantitative foundation from which the protection-operation dilemma can be further interrogated and resolved.

\section{Quantifying Internal Information Leakage}
\label{sec:quantifying}

Information lost whilst idling due to the native dynamics of a device manifests as the spreading of information that is initially localized. Information initially localized to a single qubit moves coherently into the rest of the system and is distributed non-locally. Thus the basic premise of our protocol is to monitor both a target qubit from which information may leak out, and a \textit{complementary} set of qubits into which information may have flowed. This is shown schematically in \cref{fig:schematic}\textbf{(c)} for two types of complementary qubit set: nearest-neigbour (blue) and random (orange). The basic mathematical object we use to characterize the amount of information lost is the Holevo quantity~\cite{holevo1973}: an import from quantum information theory that quantifies the amount of classical information that a channel and an ensemble of messages (an alphabet) can carry \cite{nielsen2011}. We discuss the Holevo quantity in detail in \cref{sec:holevo} for the interested reader.

The protocol in an ideal system, with no other sources of error, is discussed in detail in \cref{sec:protocol}. The protocol involves initializing a target qubit in one of two message states $|0\rangle$ and $|1\rangle$ which encode a classical bit. By calculating the Holevo quantity both over the target qubit alone $\chi^{S}$, and over a complementary set $q$ of qubits which includes the target qubit $\chi^{SQ}$, we can quantify how much extra information we get about the initial classical binary message by looking at \textit{non-local} degrees of freedom. If the difference between these two Holevo quantities is finite, then information has leaked out of the target qubit. In \cref{sec:errors} we address other sources of error, and discuss how to unravel them from true information leakage due to internal dynamics. The major issue we identify is shot noise which can artificially induce a significant signature of information loss. We address this issue in \cref{sec:experimental} by exploiting central limiting behaviour, and present an ansatz for extrapolating information loss at zero shot noise during our experimental implementation of the protocol.

We remark on two major elements of the protocol which impose constraints on the types of device to which it is applicable. The platform must be capable of (i) full state tomography on at least two qubits (a target qubit, and at least one other qubit in the device) (ii) initializing the target qubit in two definite states. We assume nothing about the microscopic features of the native physics of a device except that (iii) the dynamics are in some way local: qubits are assumed to couple more strongly to their neighbours than qubits further away in an array. This means we can look at nearest-neighbour qubits \textit{and} random qubits, both of which are equally susceptible to single-qubit errors and shot noise but are \textit{not} equally sensitive to information leakage affecting a fixed target qubit.

\subsection{The Holevo Quantity}
\label{sec:holevo}

The Holevo quantity $\chi$ over an alphabet $\mathcal{A} = \{p_k, \hat{\rho}_k\}$ and channel $\mathbf{\Lambda}$ quantifies - in bits - the accessible information that is carried by the protocol wherein states $\hat{\rho}_k \in \mathcal{H}_i$ are selected with probability $p_k$ and transmitted through the channel $\mathbf{\Lambda}[\hat{\rho}_k] \in \mathcal{H}_f$ \cite{holevo1973, nielsen2011}. The Holevo quantity $\chi$ is given by
\begin{equation}\label{eqn:holevo}
    \chi = S\left(\sum_k p_k \mathbf{\Lambda}[\hat{\rho}_k] \right) - \sum_k p_k S\Bigl(\mathbf{\Lambda}[\hat{\rho}_k]\Bigr).
\end{equation}
The Holevo quantity $\chi$ has been used to characterize information in many-body systems before, and is a natural quantity with which to investigate information loss \cite{nicokatz2022information, nicokatz2022memory, yuan2022quantum, zhuang2023dynamical}. We critically note that the initial space $\mathcal{H}_i$ and final space $\mathcal{H}_f$ are not generally identical. Consider e.g. a protocol which prepares a target qubit in a specific state, but performs measurements on the combined final state of the target qubit and its nearest-neighbours.

When maximized over all possible input alphabets $\mathcal{A}$, $\chi$ gives the maximum classical capacity of the channel $\mathbf{\Lambda}$. However, this maximization is often not possible in practice: the precise nature of a device's underlying Hamiltonian, and thus of the channel $\mathbf{\Lambda}$, is subject to debate - and the Hamiltonian parameters are subject to random fluctuations and drift. However, quantum computing platforms should approximately realize identities on idling qubits. Thus it is reliable to define an alphabet of equi-probable $p_k = 1/\text{dim}(\mathcal{H}_i)$, pure, and orthogonal messages which cover the state space $\mathcal{H}_i$ (this choice saturates $\chi$ when $\mathbf{\Lambda}$ is the identity channel).

We now discuss another important property of the Holevo quantity which becomes critically relevant when we try to account for shot noise in \cref{sec:experimental}. It is intuitively appropriate to consider $\chi$ as encapsulating the notion of \textit{distinguishability}. Direct inspection the form of $\chi$ reveals that an alphabet of pure states which are not all mutually orthogonal will cause the first term of \cref{eqn:holevo} to fail to saturate, resulting in a low value of $\chi$. Conversely, an alphabet of states which are close to maximally mixed states will saturate the first term, but also saturate the second - resulting in a low value for $\chi$ overall. Thus the Holevo quantity can be understood as a careful balancing act between the mixedness of the source (first term) and purity of the individual messages (second term) which together capture \textit{distinguishability}. This may seem like a pedantic point, but it becomes critical when accounting for the effect of shot noise. Whilst the Holevo quantity is monotonically decreasing under the implementation of a CPTP channel, shot noise can not be represented as such, and can thus artificially increase the Holevo quantity. More precisely, shot noise is realised as random contributions to elements of the final density matrices $\mathbf{\Lambda}[\hat{\rho}_k]$. These contributions can increase distinguishability but does not actually increase the amount of information that the channel can bear. This effect becomes exacerbated in the case that the dimension of the initial space is smaller than the dimension of the final space. If $\text{dim}(\mathcal{H}_i) < \text{dim}(\mathcal{H}_f)$, there are more elements in density matrices drawn from the final space, and thus shot noise introduces more differences \textit{between} output density matrices.

In the context of this article (and in the absence of shot noise, which we address separately), the Holevo quantity quantifies how much information we can access about the initial state of a qubit given access to it \textit{and} access to some other subset of qubits on the device. If no information has coherently leaked out of the target qubit, then access to some other region should not give us additional information about its initial state.

\subsection{Protocol for Quantifying Idle Information Loss}
\label{sec:protocol}

We now leverage Holevo quantities computed over two subsets of qubits on a device into protocol which quantifies the information lost to other qubits during idling. We first introduce the Holevo quantities $\chi^{S}$ and $\chi^{SQ}$, respectively computed on the final reduced density matrices (i) of the target qubit alone $\hat{\rho}_k^{S}$, and (ii) of the target qubit and a complementary set of $M-1$ qubits in the array $\hat{\rho}_k^{SQ}$. Note here that $\hat{\rho}_k^{S} = \text{Tr}_E [\hat{\rho}_k^{SQ}]$, and thus by the monotonicity of the Holevo quantity under the partial trace:
\begin{equation}
    \Delta \chi^{SQ} \coloneq \chi^{SQ} - \chi^{S} \geq 0
\end{equation}
with equality only when information is fully localized to the target qubit, or when $\chi^{SQ} = \chi^{S} = 0$ and information has completely left the final space. Since the Holevo quantity yields the average number of bits of information that can be transmitted by messages passed through the channel, $\Delta \chi^{SQ}$ can also be interpreted as the extra bits of information we can access about the initial state of the target qubit $S$ given access to the complementary qubits $Q$. In ideal systems without shot noise or other sources of error $\Delta \chi^{SQ}$ suffices to identify and quantify information loss due to information leakage. Finite $\Delta \chi^{SQ}$ indicates that some information has left the target qubit in a coherent fashion, and is distributed non-locally: i.e. that the device can't `do nothing' perfectly, even in ideal conditions. The protocol in full is as follows:
\begin{enumerate}
    \item Initialize every qubit except the target qubit $S$ in an arbitrary state $\hat{\rho}_E$ (we take pure separable product states of the logical single-qubit states throughout).
    \item Initialize the target qubit $S$ in the state $|0\rangle$.
    \item Wait for a fixed period of time $T$ (we measure immediately to address the `best case' scenario with minimal information leakage such that $T$ is the readout time of the device) \footnote{$T$ is arbitrary in principle, and can be tuned to address different regimes or properties of a device. Taking $T$ to be e.g. fifty two-qubit gate times could give us a good indicator of the \textit{total} information leakage during a complete computation.}.
    \item Perform full state tomography on the combined state $\hat{\rho}^{SQ}_0$ of the target and complementary qubits (the subscript denotes that this state is conditioned upon the initial state, prepared in step 2).
    \item Repeat from step 1, with the target qubit $S$ initialized in the state $|1\rangle$, generating the state $\hat{\rho}^{SQ}_1$.
    \item Process the states into a sample value for $\Delta \chi^{SQ}$.
    \item Check the condition $\Delta \chi^{SQ} > 0$.
    \item (Optional, to address noise) Repeat from step 1 for a large number of samples to build up statistics for $\Delta \chi^{SQ}$.
\end{enumerate}

\begin{figure}[ht]
        \includegraphics[width=\linewidth]{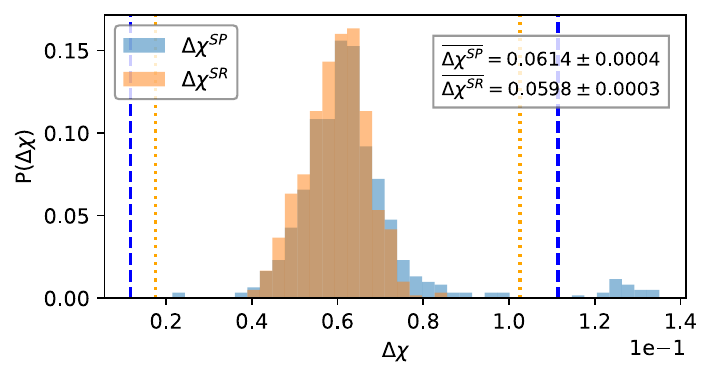}
        \caption{\textbf{Experimental sample statistics for $\Delta \chi$} Statistics of $\Delta \chi^{SP}$ and $\Delta \chi^{SR}$ for nearest-neighbour $P$ (blue) and random $R$ (orange) complementary sets of qubits respectively. Shot count is $N_S = 4000$. $\Delta \chi$ quantifies how much extra information about the target qubit can be accessed by measuring non-target qubits in the complementary sets $P$ or $R$, and can be artificially increased by shot noise, which should have an identical effect on both distributions. The ranges determined by \cref{eqn:filter} at a tolerance of $K=4$ are shown as blue dashed lines for $\Delta \chi^{SP}$, and orange dotted lines for $\Delta \chi^{SR}$. Samples which fall outside of this range are classified as outliers, and excluded from the numerical analysis. Inset shows means and standard errors for the non-outlier region of each distribution; evidencing that \cref{eqn:qubic-inequality-average} is satisfied and idle information loss has been directly detected after accounting for shot noise. A noticable bump of outliers at larger values of \textit{only} $\Delta \chi^{SP}$ is evidence of `bad qubits' that have strongly coupled to their neighbours and are leaking lots of information during idling. Statistics are comprised of 1209 total samples, taken across four months (Dec 2023 - Mar 2024) from all the Falcon 5.11 series of devices.}

        \label{fig:fig1}
\end{figure}

\subsection{Unravelling Information Leakage from Other Sources of Error}
\label{sec:errors}

Unravelling information leakage from other sources is, in principle, a serious operational issue. As we discussed in \cref{sec:holevo}, shot noise may artificially increase distinguishability. Moreover, other sources of error may artificially increase the difference $\Delta\chi^{SQ}$, which is not subject to the same monotonicity conditions as the Holevo quantity alone. In essence, $\Delta\chi^{SQ}$ can be non-zero in realistic systems \textit{even} if information is fully localized to the target qubit. To deal with this we benchmark our results on nearest-neighbour qubits $Q=P$ against random (non nearest-neighbour) qubits in the array $Q=R$, whilst keeping the target qubit $S$ fixed. Both of these sets of complementary qubits should, when averaged over a large number of samples - be equally affected by e.g. shot noise, SPAM errors, and environmentally-induced decoherence. However, the randomly selected qubits should bear much less information about the initial state of $S$; and thus they can serve as a benchmark for all other sources of error in the device. Nevertheless, we choose to carry out our experiments at very short timescales, two orders of magnitude less than the dephasing times of the devices we use, to mitigate the effect of the environment as much as possible. Our protocol does not make use of any gates in its implementation, and thus gate errors are vacuously irrelevant. We implement these benchmarking procedures in \cref{sec:experimental} and ultimately find that shot noise is the dominant contribution to $\Delta\chi$ in the system of interest. We suggest, but can not prove, that channels comprised of a linear combination of single-qubit processes can not systematically increase $\Delta\chi^{SQ}$; as such single-qubit processes can not systematically transfer information from $S$ to $Q$.

We first invoke the assumption that the underlying dynamics are, in a sense, local, and that the idea of `nearest-neighbours' is thus well-defined. In such systems, information should flow from the target qubit to its nearest-neighbours, and then through them to more distant qubits in the array. For a fixed target qubit $S$, we can then define two types of complementary sets of qubits: $P$ corresponding to the plaquette of the $M-1$ nearest-neighbours of $S$, and $R$ corresponding to the target qubit and $M-1$ randomly selected (excluding nearest-neighbour) qubits in the array. Since $P$ and $R$ both contain the same number of qubits they should, after averaging, be equally susceptible to single-qubit errors and shot noise (see \cref{sec:tomography} for a detailed discussion on the origin of this shot noise in IBM devices). In an ideal system without other errors or shot noise $\Delta \chi^{SP} - \Delta \chi^{SR} > 0$ should be a strict inequality (except at very late times where the information becomes fully delocalized). Individual samples may violate this inequality, but since both $P$ and $R$ are equally susceptible to all other sources of noise, these effects should cancel out after averaging over a large number of samples. We thus relax this condition to a statistical inequality
\begin{equation}\label{eqn:qubic-inequality-average}
    \overline{\Delta\chi^{SP}} - \overline{\Delta\chi^{SR}} > 0
\end{equation}
that should hold in the presence of non-trivial information leakage. Moreover, \cref{eqn:qubic-inequality-average} can be easily realized as the alternative hypothesis $H_1$ for significance testing; which we do in \cref{sec:experimental}. 

This inequality can be extended more formally by suggesting an ansatz form for $\Delta\chi$ as follows:
\begin{equation}\label{eqn:generic-ansatz}
    \Delta \chi^{SQ} = f\left(\eta^{SQ}, \eta^{SQ}_\text{SHOTS}, N_\text{S}\right).
\end{equation}
Which is a function of the excess accessible information $\eta^{SQ}$, artificial information due to shot noise $\eta^{SQ}_\text{SHOTS}$, and the shot count $N_\text{S}$. We can aggregate statistics at different shot counts, perform a scaling analysis in $N_\text{S}$, and extrapolate $\eta^{SQ}$ as $N_\text{S} \to \infty$. We develop just such an ansatz based on our experimental data in \cref{sec:experimental}. This also formalizes the intuitive notion of both the sets $P$ and $S$ of complementary qubits being `equally susceptible' shot noise. We say that both are `equally susceptible' to shot noise if $\eta^{SP}_\text{SHOTS} \approx \eta^{SR}_\text{SHOTS}$ after fitting. With shot noise mitigated, the difference $\eta^{SP}-\eta^{SR}$ then gives us the amount of information which has leaked from the target qubit into its neighbours.

In actual experiment (see \cref{sec:experimental}), we find $\eta^{SR} \to 0$ exactly, supporting our previous assertion that other single-qubit sources of error can not systematically transfer information to distant regions of the device. This indicates that $\eta^{SP}$ alone suffices to determine the amount of idle information loss in future experiments.

\section{Experimental Results and a Quantification of Idle Information Loss on IBM Devices}
\label{sec:experimental}

The experimental implementation of our protocol was carried out on the 27-qubit Falcon 5.11 series of IBM's quantum computing devices; the architecture of such devices is shown in \cref{fig:schematic}\textbf{(c)}. We incorporate two additional steps into the general protocol given in \cref{sec:protocol}: (i) we only select target qubits with the highest possible coordination number $n_c = 3$, which should yield the strongest signatures of idle information loss. And (ii) in step 1 of the protocol, we also simultaneously randomize over target qubits subject to the coordination constraint, and the Falcon 5.11 devices \texttt{ibm\_algiers}, \texttt{ibm\_cairo}, \texttt{ibm\_hanoi}, and \texttt{ibmq\_kolkata} themselves. This corresponds to the generation of statistics for $\Delta\chi$ which are device agnostic - i.e. the user is interested only in running their computation job on a 27-qubit Falcon 5.11 device, and doesn't care about which specific device to which their jobs are assigned. We performed readout immediately after state preparation, such that the wait time in step 3 of the protocol was just the readout time of the given device (between 700ns and 900ns). This minimizes the effect of the environment, as these readout times are several orders of magnitude lower than typical T1 and T2 times. This also means that our results represent the best case scenario, and can not be improved by e.g. dynamical decoupling, as such processes are not possible during readout. Due to the existence of shot noise, the resulting density matrices after state tomography can have negative eigenvalues. We use a maximum-likelihood reconstruction method to rephysicalize the density matrices before post-processing \cite{smolin2012efficient}. A detailed discussion of the tomographic process is given in \cref{sec:tomography}, and the maximum-likelihood reconstruction of aphysical density matrices is discussed in \cref{sec:aphysicality}.

As discussed in \cref{sec:protocol}, we consider both nearest-neigbour $P$ and random $R$ sets of complementary qubits for each target qubit, and carry out separate collections of experiments for each at a range of different shot counts $N_S$. The number of samples for all shot counts $N_S$ on both nearest-neighbour plaquettes and random qubits is summarized in \cref{tab:samples} in \cref{sec:tomography}. All in all, our analysis of the Falcon 5.11 series involves the results of over 3500 experiments taken across four months (Dec 2023 - Mar 2024), and represents a broad-spectrum comprehensive investigation of idle information loss on these devices. As the runtime increases proportionally with the shot count, most of this time was spent collating results for higher shot count samples; thus a direct comparison where statistics are compared for lots of low shot count samples, is likely the best approach for characterizing idle information loss on IBM devices in the future.

\subsection{Results and Analysis}
\label{sec:results}

\begin{figure}[ht]
        \includegraphics[width=\linewidth]{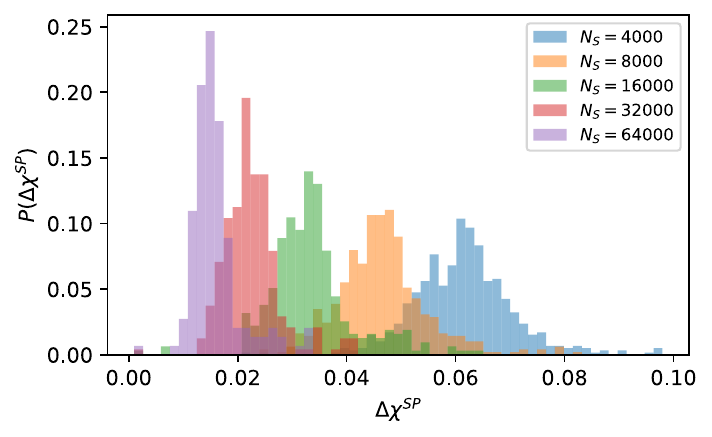}
        \caption{\textbf{Statistics of $\mathbf{\Delta\chi^{SP}}$ for a range of shot counts} Outlier-excluded $K=4$ statistics of $\Delta\chi^{SP}$ for a range of shot counts $N_S$. As shot count is doubled, the mean value of each set of statistics appears to scale by a factor of $1/\sqrt{2}$, indicating central limiting behaviour in $N_S$ which dominates signatures of idle information loss. A total of $1849$ samples are distributed across the five different values of $N_S$, a breakdown of which is provided in \cref{tab:samples}. }
        \label{fig:fig2}
\end{figure}

The experimentally determined statistics of $\Delta \chi^{SP}$ and $\Delta \chi^{SR}$ for 1209 total samples across all devices are shown in \cref{fig:fig1}. The shot count for both was fixed at the standard value of $N_S = 4000$. Both take a normally distributed form within certain limits, with $\Delta \chi^{SP}$ exhibiting an additional small bump in the range $0.12-0.14$. Were this bump due to any of the other processes discussed: decoherence, SPAM errors, shot noise, we would expect bumps to also appear in the statistics of $\Delta \chi^{SR}$. As this bump is not present in the statistics of $\Delta \chi^{SR}$, it stands to reason that this is strictly due to some target qubits coupling strongly to their nearest-neighbours. Moreover, we identify a bump in the statistics of $\Delta \chi^{SP}$ at larger shot counts \textit{at the exact same position} (see \cref{sec:additional}), evidencing a true signature of idle information loss. The samples that form this bump can be justifiably interpreted as `bad qubits' which have hybridized with their nearest-neighbours to such an extent that information leakage is dominating other sources of error. Over $10\%$ of the information about the initial state of such a qubit is stored non-locally.

A more subtle feature of \cref{fig:fig1} is that the statistics of $\Delta \chi^{SP}$ show a thicker tail at larger values than $\Delta \chi^{SR}$, suggesting that the inequality \cref{eqn:qubic-inequality-average} is satisfied even when these bad qubits are excluded. To quantify this difference, we first filter the statistics for outliers by simple box-filtering according to the condition \cite{schwertman2004simple, yang2019outlier},
\begin{equation}\label{eqn:filter}
    \text{Q1} - K \times  \text{IQR} \leq \Delta \chi^{SQ}_s \leq  \text{Q3} + K \times \text{IQR}
\end{equation}
where we have introduced the subscript $s$ to denote individual samples of $\Delta \chi^{SQ}$. Q1 and Q3 are the first and third quartiles of the full sample statistics  $\Delta \chi^{SQ}$ respectively, and IQR is their inter-quartile range. Conservative filtration typically takes $K=1.5$, whilst $R \to \infty$ corresponds to no filtering at all. We take $K=4$ throughout as this both reliably contains the large Gaussian part of the distributions, and also excludes the bad qubit bump in the statistics of $\Delta \chi^{SP}$. We superimpose the boundaries of the box defined by \cref{eqn:filter} on \cref{fig:fig1} as dashed blue lines for $\Delta \chi^{SP}$ and a dotted orange line for $\Delta \chi^{SR}$ \footnote{It is thus interesting to point out that we can thus invoke $K$ as a tuning parameter which defines precisely what is meant by a `bad qubit'. The part of the distribution to the right of the box defined by $K$ can inform us of the probability that any given qubit will be a `bad' one up to a certain informational tolerance. In cases where the entire device is required to complete a computation, or bad qubits are not identified and excluded before runtime, this probability could be used to place bounds on ultimate ruin: wherein said computation fails completely. We defer this topic to future study.}. After filtering, we can compute the means $\overline{\Delta \chi^{SQ}}$ and standard errors in the means $\sigma^{SQ}$ of the resulting datasets, which are displayed in the inset of \cref{fig:fig1}. The results satisfy \cref{eqn:qubic-inequality-average}, and provide a clear `smoking gun' for idle information loss in the investigated devices. More formally, we carried out a one-tailed Welch's t-test with null hypothesis $H_0 : \overline{\Delta \chi^{SP}} \leq \overline{\Delta \chi^{SR}}$ and alternative hypothesis $H_1 : $ \cref{eqn:qubic-inequality-average}. This yields a $z$-value of $z=3.30$ (corresponding to a $p$-value of $p=0.00049$) which implies that the inequality holds to a high degree of statistical significance. We perform similar tests at a range of shot counts $N_S$, the resulting $z$-values and corresponding $p$-values of which are summarized in \cref{tab:welchfull}. These results indicate that $H_0$ can be rejected to a very high degree of statistical significance, and thus that true signatures of idle information loss have been detected, across all investigated shot counts. 

\begin{table}[ht]
    \centering
    \renewcommand{\arraystretch}{1.2}
    \begin{tabular}{|c|c|c|} \hline
         \textbf{Shot Count} $N_S$ & $\mathbf{z}$\textbf{-value} & $\mathbf{p}$\textbf{-value} \\ \hline\hline
         4000 & $z = 3.3024$ & $p = 4.9448 \times 10^{-4}$ \\ \hline
         8000 & $z = 6.0996$ & $p = 7.8796 \times 10^{-10}$ \\ \hline
         16000 & $z = 6.7304$ & $p = 2.3839 \times 10^{-11}$ \\ \hline
         32000 & $z = 5.4390$ & $p = 4.4893 \times 10^{-8}$ \\ \hline
         64000 & $z = 5.0447$ & $p = 4.8972 \times 10^{-7}$ \\
         \hline
    \end{tabular}
    \caption{\textbf{Hypothesis testing for signatures of idle information loss} Resulting $z$-values and corresponding $p$-values of single-tailed Welch's t-tests given the null hypothesis $H_0 : \overline{\Delta\chi^{SP}} \leq \overline{\Delta\chi^{SR}}$ after $K=4$ filtering, for all shot counts $N_S$. }
    \label{tab:welchfull}
\end{table}

\begin{figure}[ht]
        \includegraphics[width=\linewidth]{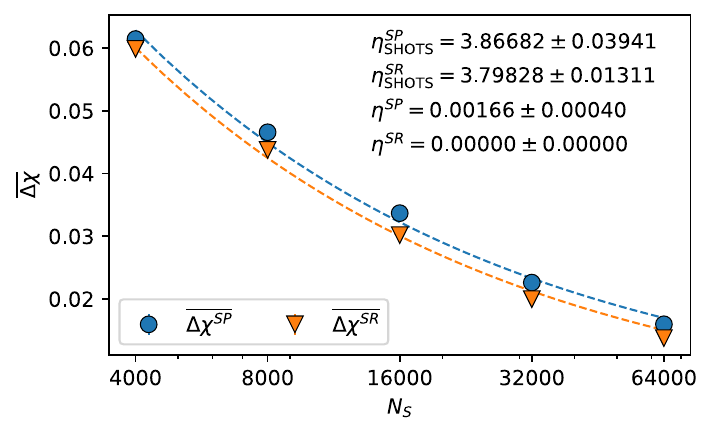}
        \caption{\textbf{Scaling analysis and zero-noise extrapolation} Outlier-filtered $K=4$ mean values $\overline{\Delta\chi^{SP}}$ and $\overline{\Delta\chi^{SR}}$ at different shot counts $N_S$; error bars shown where visible. Dashed lines show a fit to these values according to \cref{eqn:chaosfit}, with inset text showing the fitted values. The extrapolated $N_S \to \infty$ zero shot noise values of $\Delta \chi^{(SQ)} \to \eta^{SQ}$ quantify non-trivial leakage of information into nearest neighbours $\eta^{SP} = 0.00166 \pm 0.00040$, and none into random distant qubits in the array $\eta^{SR} = 0$. Additionally, both complementary sets of qubits $P$ and $R$ seem equally affected by shot noise $\eta_\text{SHOT}^{SP} \approx \eta_\text{SHOT}^{SR}$. A total of $3591$ samples are distributed across the five different values of $N_S$ and two types of complementary qubit sets, a breakdown of which is provided in \cref{tab:samples}. }
        \label{fig:fig3}
\end{figure}

\subsection{Scaling analysis}
\label{sec:scaling}

We now interrogate the effect of shot noise more rigorously by investigating how the statistics and mean values change as we vary the number of shots $N_S$. This will form the basis of developing an ansatz of the form of \cref{eqn:generic-ansatz} with which a scaling analysis can be carried out. The results of this investigation are shown in \cref{fig:fig2} which shows the statistics of $\Delta \chi^{SP}$ after $K=4$ filtering as a function of shot count $N_S$. The resulting distributions still appear normally distributed, but drift to lower mean values with smaller variances as $N_S$ is increased. Interestingly, the mean value of $\overline{\Delta \chi^{SP}}$ falls by a factor of $1/\sqrt{2}$ every time $N_S$ is doubled. This is further evidence that shot noise, which should follow central limiting behaviour and scale as $1/\sqrt{N_S}$, is completely dominating the effects of idle information loss. We formalize this intuition by suggesting the simple ansatz
\begin{equation}
\label{eqn:chaosfit}
\overline{\Delta \chi^{SQ}} = \eta^{SQ} + \frac{\eta^{SQ}_\text{SHOTS}}{\sqrt{N_S}}
\end{equation}
which incorporates both a flat information leakage $\eta^{SQ}$ and a term which captures the distinguishability introduced by shot noise $\eta_\text{SHOTS}$. Essentially, $\eta^{SQ}$ can be interpreted as the number of additional bits of information we can retrieve about the initial state of the target qubit given access to the other qubits in the complementary set $Q$ in the zero shot noise limit $N_S \to \infty$.

We present a fit of the mean values of the $\Delta\chi^{SP}$ distributions (after $K=4$ filtering) of \cref{fig:fig2} to the ansatz of \cref{eqn:chaosfit} in \cref{fig:fig3}, where we find very good agreement between the data and our ansatz. We also present a fitting for $K=4$ filtered $\Delta\chi^{SR}$ statistics as a benchmark. The fitting procedure allows us to extrapolate values for  $\eta^{SQ}$ and $\eta^{SQ}_\text{SHOTS}$, which are shown for both $SP$ and $SQ$ in the inset text of \cref{fig:fig3}. Standard errors on these extrapolated values are determined by a simple bootstrapping scheme in which each of the data points are randomly drawn from normal distributions determined by their respective means and standard deviations. These extrapolated values reveal that, as expected, both nearest-neighbour plaquettes and randomly selected qubits are equally affected by shot noise $\eta_\text{SHOTS}^{SP} \approx \eta_\text{SHOTS}^{SR}$. However, the underlying idle information loss saturates to a low but finite value $\eta^{SP} = 0.00166\pm 0.00040$ for nearest-neighbour plaquettes, and exactly to zero (to five decimal places) $\eta^{SR} = 0$ for randomly selected qubits. This evidences the conjecture made in \cref{sec:errors} that the single-qubit processes that lead to e.g. SPAM errors and decoherence have no systematic effect on $\Delta\chi^{SQ}$, and that we can treat $\eta^{SQ}$ as a true zero shot noise quantification of idle information leakage. Ultimately, the excess information gained by having access to the joint state $SP$ when compared to $SR$ yields, on average, an additional $0.00166$ bits of information about the target qubit $S$. This is a low value, but it represents a direct quantification of the impact of idle information loss, and a \textit{fundamental} limit on how well qubits in the Falcon 5.11 series of devices can perform.

\section{Conclusions}

The foremost result of this article is a protocol exploiting the Holevo quantity from quantum information theory. This protocol provides a flexible, scalable, device non-specific solution to the burgeoning problem of quantifying idle information leakage in quantum computing platforms. Sufficient degradation in single-qubit protection could destroy the ability of a device to actually carry out quantum computations as information propagates, and our protocol serves as a direct quantification of the information lost (in bits) to this effect. A central component of our protocol is simply waiting, i.e. implementing the empty circuit; and it can thus be easily run during downtime with minimal oversight, replacing otherwise wasted time with a valuable characterization of errors in the computing platform.

The secondary result of this article is the experimental implementation of our protocol on 3500 samples carried out across four months on all four of IBM's Falcon 5.11 series of devices. The results of this analysis reveal (i) that a measurable amount of information about the state of any given qubit is leaking out during idling and (ii) the existence of `bad qubits' which leak over $10\%$ of a classical bit of information into their immediate surroundings. The `smoking gun' of idle information loss takes the form of a statistical inequality which accounts for the effects of e.g. SPAM errors, decoherence, and shot noise. We find this inequality to be satisfied to remarkably high degrees of statistical significance at all shot counts. We also determine an ansatz from which exact idle information loss at zero shot noise can be extrapolated. The results of this extrapolation indicate that, after filtering for bad qubits, approximately $0.2\%$ of the information stored locally is lost to idle information loss during a single readout time. This is a low, and but crucially non-zero, value which represents a \textit{fundamental} limit on how well the Falcon 5.11 series of devices can perform.

Overall, our results indicate that, in contrast to what has been suggested in literature \cite{berke2022transmon}, unwanted many-body effects are not a significant concern in current-generation IBM devices when compared to shot noise; with the exception of occasional bad qubits. However, the main finding of our work is that a measurable albeit small amount of information is \textit{already} being lost in these systems and we can study this accurately with our method. This should allow us to systematically understand the impact of innovating technologies on the protection-operation dilemma. The near future of quantum computing promises dramatic scale-ups of system sizes, nascent error correcting hardware, novel approaches to localizing information, and fledgling fault-tolerance. Our work provides a flexible, powerful, scalable protocol to quantify idle information loss in all these settings. This represents a decisive step towards addressing the threat many-body effects pose to high-fidelity idling, and thus to the long-term large-scale stability of generic quantum computing platforms.

\section{Acknowledgements}
\label{sec:acknowledgements} 

A.N.-K. would like to thank S. Bose, C. Turner, C. Berke, and O. Dial for fruitful discussions at preliminary research stages. A.N.-K. sincerely thanks S. Shaikh and C. Bleger for their support. N.K. would like to give thanks to the QuSys group for discussions throughout the project, and also to G. García-Pérez, S. Filippov, and E. Borrelli for giving feedback and useful references. J.G. is supported by a SFI- Royal Society University Research Fellowship and is grateful to IBM Ireland and Microsoft Ireland for generous financial support. 

\bibliography{refs}

\appendix

\section{Additional Results}
\label{sec:additional}

\begin{figure*}
\centering
\subfloat[$N_S$ = 8000]{\centering
        \includegraphics[width=.48\linewidth]{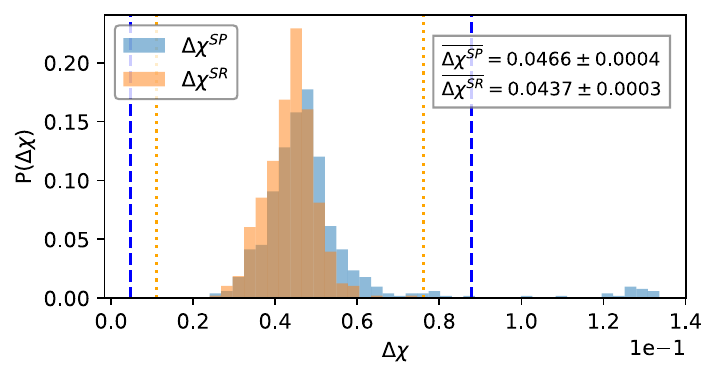}
        \label{fig:additional-results-8000}
}
\hfill
\subfloat[$N_S$ = 16000]{\centering
        \includegraphics[width=.48\linewidth]{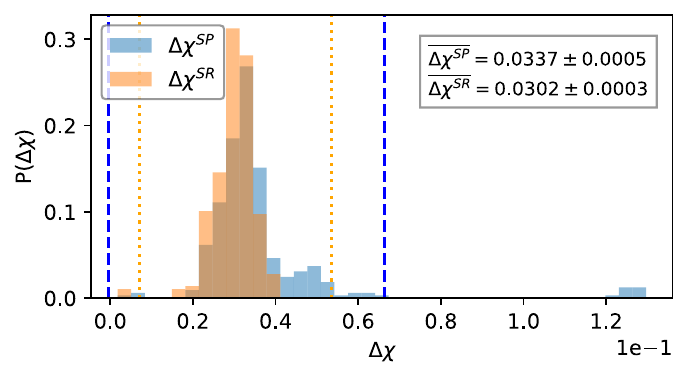}
        \label{fig:additional-results-16000}
}

\subfloat[$N_S$ = 32000]{\centering
        \includegraphics[width=.48\linewidth]{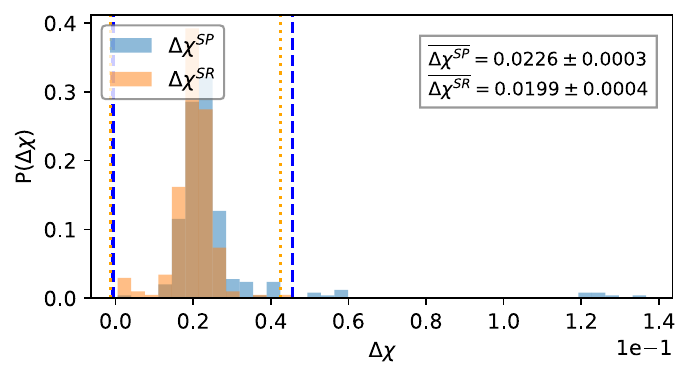}
        \label{fig:additional-results-32000}
}
\hfill
\subfloat[$N_S$ = 64000]{\centering
        \includegraphics[width=.48\linewidth]{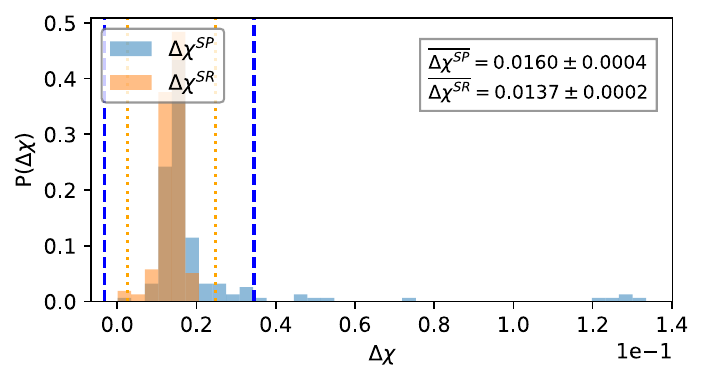}
        \label{fig:additional-results-64000}
}
\caption{\textbf{Experimental sample statistics for $\Delta \chi$} Statistics of $\Delta \chi^{SP}$ and $\Delta \chi^{SR}$ for nearest-neighbour $P$ (blue) and random $R$ (orange) complementary sets of qubits respectively. Different shot counts $N_S$ shown in \textbf{(a)} $N_S = 8000$, \textbf{(b)} $N_S = 16000$, \textbf{(c)} $N_S = 32000$, \textbf{(d)} $N_S = 64000$. The ranges determined by \cref{eqn:filter} at a tolerance of $K=4$ are shown as blue dashed lines for $\Delta \chi^{SP}$, and orange dotted lines for $\Delta \chi^{SR}$. Inset shows means and standard errors for the non-outlier region of each distribution. A noticable bump of outliers at larger values of \textit{only} $\Delta \chi^{SP}$ is visible in all cases. Statistics are comprised of 2369 total samples, taken across four months (Dec 2023 - Mar 2024) from all the Falcon 5.11 series of devices.}
\label{fig:additional-results}
\end{figure*}

Here we present additional results in \cref{fig:additional-results} for the statistics of $\Delta\chi^{SP}$ and $\Delta\chi^{SR}$, similar to \cref{fig:fig1} in the main text at different shot counts $N_S$. Dashed blue and dotted orange lines indicate the $K=4$ box-filtration fences according to the outlier filtering of \cref{eqn:filter} in the main text for $\Delta\chi^{SP}$ and $\Delta\chi^{SR}$ respectively. In all cases we identify a bump in $P(\Delta\chi^{SP})$ at $\Delta\chi^{SP} \gtrsim 0.12$, for the nearest-neighbour complementary set of qubits $P$ \textit{only}. As discussed in the main text, this bump corresponds to `bad qubits' which leak considerable information (over 10\% of a classical bit) into their surroundings. The fact that this bump is only visible in the nearest-neighbour complementary set data $P$, and that it doesn't change position with changing shot count, suggests that it is entirely independent of shot noise or other sources of error and is in fact a direct quantification of idle information loss.

We also note that, as the shout count is doubled between the panels \cref{fig:additional-results}\textbf{(a)}-\textbf{(d)}, the centre of the main Gaussian distribution decreases; indicating that much of the signal is dominated by shot noise. We evaluate the mean values $\overline{\Delta\chi}$ and their respective standard errors in the mean, both of which are shown in the inset boxes in each panel. The results of Welch's t-tests identical to the one carried out in the main text are summarized in \cref{tab:welchfull}, indicating that the statistics for $\overline{\Delta\chi^{SP}}$ and $\overline{\Delta\chi^{SR}}$ are different to a high degree of statistical significance: with $z$-values in excess of $z=5$ for all the results shown in \cref{fig:additional-results}. This indicates a positive detection of idle information loss at all shot counts investigated; even if the absolute value of this idle information loss is very small.

Finally, we note that at the higher shout counts $N_S = 32000$ and $N_S=64000$ shown in \cref{fig:additional-results}\textbf{(c)} and \cref{fig:additional-results}\textbf{(d)} respectively, there is a finite probability of locating some results outside of the filter boundary but well below the threshold of $\Delta\chi \gtrsim 0.12$ for unambiguous `bad qubits'. This corresponds to qubits which are leaking enough information to be visible above the background of shot noise but which may not have failed completely.

\section{Details of State Tomography}
\label{sec:tomography}

State tomography in IBM devices is carried out by measuring all $M$ qubits in the output register ${j_1,j_2,\cdots,j_M}$, where $j_r$ indexes the physical location of the $r$-th output qubit, simultaneously in the three Pauli bases. Taking the standard Pauli matrices, $\sigma^0 = \mathbb{I}$, $\sigma^1 = X$, $\sigma^2 = Y$,  $\sigma^3 = Z$, the statistics of each Pauli string $P_b = \sigma^{b_1}_{j_1}\sigma^{b_2}_{j_2}\cdots\sigma^{b_M}_{j_M}$ with $M$ non-identity elements ($b_r \neq 0$ such that there are $3^M$ of this family of string in total) is determined by these measurements. The result of each measurement is a bitstring, and the result of a large number $N_S$ of measurements - called `shots' - is a dictionary $D_b$ of bitstrings. Taking $N_S$ sufficiently large ensures that the sample statistics of measuring a specific bitstring given the Pauli string $P_b$ are close to the population statistics. These statistics for the reduced space of Pauli strings can then be aggregated into marginal values which yield statistics for the full space of $4^M$ Pauli strings. This is done by simply aggregating shots from different Pauli strings together wherever they coincide everywhere except where identities occur in the desired marginal. 

\begin{table}[ht]
    \centering
    \renewcommand{\arraystretch}{1.1}
    \begin{tabular}{|c|c|c|} \hline
         \textbf{Shot Number} $N_S$ & \textbf{Complementary Set} & \textbf{Samples} \\ \hline\hline
         \multirow{2}{*}{4000} & $P$ & 609 \\ & $R$ & 600 \\ \hline
         \multirow{2}{*}{8000} & $P$ & 507 \\ & $R$ & 480 \\ \hline
         \multirow{2}{*}{16000} & $P$ & 324 \\ & $R$ & 288 \\ \hline
         \multirow{2}{*}{32000} & $P$ & 252 \\ & $R$ & 204 \\ \hline
         \multirow{2}{*}{64000} & $P$ & 157 \\ & $R$ & 157 \\
         \hline
    \end{tabular}
    \caption{\textbf{Total sample counts for all experimental implementations} Sample counts, before filtering according to \cref{eqn:filter}, for each shot count and type of tomographic complementary set: nearest-neigbour $P$ and random (excluding nearest-neighbour) $R$. Each sample corresponds to a single realization of steps 1-7 of the protocol discussed in \cref{sec:protocol}.}
    \label{tab:samples}
\end{table}

As a concrete example, consider an output register of three qubits, and lets say we are interested in the value of $\langle X I Y \rangle$. This is given by the statistics of $X X Y = P_1$, $X Y Y = P_2$, and $ X Z Y = P_3$. For each of these $P_b$ statistics, there is a corresponding dictionary of results $D_b = \{B_b^{s}\}$ where the $B_b^s$ are bitstrings $x$ which are simply the results of any single shot (measurement) given $b$. The statistics of $P_1$ are given by calculation of the probability distribution $P_b(B_b = x) = P(B = x | b)$. An example dictionary for $P_1$ for $N_S = 10$ shots might be
\begin{equation}
    D_1 = \{101,101,101,111,001,101,101,001,101,100\}
\end{equation}
with an associated probability distribution
\begin{equation}
    P_1(B_1 = x) = \begin{cases}
               0.6 \quad \text{if}~x = 101 \\
               0.1 \quad \text{if}~x = 111 \\
               0.2 \quad \text{if}~x = 001 \\
               0.1 \quad \text{if}~x = 100 \\
               0.0 \quad \text{otherwise}
            \end{cases}.
\end{equation}
Now consider the following example dictionaries for $P_2$ and $P_3$:
\begin{align}
    D_2 &= \{111,101,111,111,011,111,101,011,101,101\}\\
    D_3 &= \{101,111,011,111,011,110,111,001,101,110\}.
\end{align}
We then aggregate the dictionaries $D_b$ into a single new dictionary $\widetilde{D}$ which describes the statistics of $X I Y$ by simply excluding the central bit of each bitstring and aggregating the dictionaries:
\begin{align}
    D_1 &\to \{11,11,11,11,01,11,11,01,11,10\}\\
    D_2 &\to \{11,11,11,11,01,11,11,01,11,11\}\\
    D_3 &\to \{11,11,01,11,01,10,11,01,11,10\}\\
    \widetilde{D} &= D_1 + D_2 + D_3
\end{align}
where $\widetilde{D}$ contains 30 elements. The statistics $\widetilde{P}(\widetilde{B} = \widetilde{x})$ of $\widetilde{D}$ are then calculated as
\begin{equation}
    \widetilde{P}(\widetilde{B} = \widetilde{x}) = \begin{cases}
               0.6\dot{6} \quad \text{if}~x = 11 \\
               0.2\dot{3} \quad \text{if}~x = 01 \\
               0.10 \quad \text{if}~x = 10 \\
               0.00 \quad \text{otherwise}
            \end{cases}.
\end{equation}
We can now evaluate $\langle X I Y \rangle$ explicitly by summing up contributions to the expectation value $1 \to 1$, $0 \to -1$ as follows
\begin{equation}
    \langle X I Y \rangle = 0.6\dot{6} (-1 \times -1) + 0.2\dot{3} (1 \times -1) + 0.1 (-1 \times 0) = 0.\dot{3}
\end{equation}
which completes our example. 

This may seem a laborious process, but it allows us to extrapolate $\mathcal{O}(4^M)$ elements of a given state's density matrix using only $\mathcal{O}(N_S 3^M)$ measurements. By decomposing the state's density matrix into a sum of $4^M$ Pauli strings (including identity elements),
\begin{equation}\label{eqn:density-matrix-construction}
    \hat{\rho} = \frac{1}{\mathcal{Z}} \sum_b \langle P_b \rangle P_b
\end{equation}
where $\mathcal{Z}$ is an appropriate normalization factor, we can readily reconstruct the quantum state of the output register using the dictionaries $D_k$. These dictionaries $D_k$ are ultimately what IBM's quantum computers return to their users. Where we discuss shot count $N_S$ in the main text, it simply refers the size of these returned dictionaries; where larger dictionaries more accurately yield the statistics of the actual population. The total number of samples of $\Delta \chi$ for both complementary qubit sets discussed in the main text (see \cref{sec:quantifying}) is shown in \cref{tab:samples} for all shot counts $N_S$ we consider.

\section{Aphysicality and Maximum-Likelihood Reconstruction}
\label{sec:aphysicality}

As discussed in \cref{sec:experimental} of the main text, shot noise due to finite $N_S$ can result in aphysical density matrices by introducing negative eigenvalues into their spectra. We correct for this using the maximum-likelihood reconstruction of the density matrix.

The tomographic process discussed in \cref{sec:tomography} yields a density matrix $\hat{\mu}$ with matrix elements $\mu_{ij}$ which is definitionally of trace unity and hermitian by inspection of \cref{eqn:density-matrix-construction}. The eigenvalues $\mu_j$ of $\hat{\mu}$ can, however, be negative; and thus $\hat{\mu}$ is generally unphysical.

We follow and briefly summarize here the maximum-likelihood mixed state reconstruction algorithm given in the work of Smolin et al. \cite{smolin2012efficient}. First we invoke the existence of some density matrix $\hat{\rho} \in \mathcal{P}$ with matrix elements $\rho_{ij}$ and eigenvalues $\rho_j$ which minimizes the 2-norm,
\begin{equation}\label{eqn:ml-algo-0}
    \text{min}_{\hat{\rho} \in \mathcal{P}}\parallel \hat{\rho} - \hat{\mu} \parallel_2^2 = \text{min}_{\hat{\rho} \in \mathcal{P}} \sum_{ij} | \rho_{ij} - \mu_{ij} |^2
\end{equation}
where $\mathcal{P}$ is the space of physical density matrices (unit trace, positive semi-definite, hermitian matrices). We remark here that \cref{eqn:ml-algo-0} is invariant under change of basis, and hence we choose to work in the eigenbasis of $\hat{\mu}$ with eigenvectors $|\mu_j\rangle$ such that,
\begin{equation}\label{eqn:ml-algo-1}
    \parallel \hat{\rho} - \hat{\mu} \parallel_2^2 = \sum_{ij} | \rho_{ij} - \mu_{j}\delta_{ij} |^2
\end{equation}
where $\delta_{ij}$ is the Kronecker delta. Clearly, \cref{eqn:ml-algo-1} is minimized when $\hat{\rho}$ is also diagonal in this basis, i.e. the eigenvectors of $\hat{\rho}$ are also $|\mu_j\rangle$, as any non-zero off-diagonal terms $\rho_{ij}$ for $i \neq j$ strictly increases the value of \cref{eqn:ml-algo-1}. This reduces the minimization procedure down from an $\mathcal{O}(\Omega^2)$ problem (where $\Omega$ is the total dimension of the system) to a minimization problem in the $\Omega-1$ eigenvalues of $\hat{\rho}$,
\begin{equation}\label{eqn:ml-algo-2}
     \text{min}_{\hat{\rho} \in \mathcal{P}} \parallel \hat{\rho} - \hat{\mu} \parallel_2^2 = \text{min}_{\{\rho_j\}} \sum_{j} | \rho_{j} - \mu_{j} |^2
\end{equation}
subject to only two constraints: that $\rho_j \geq 0$, and that $\sum_j \rho_j = 1$. The reconstructed density matrix is finally given by
\begin{equation}
    \hat{\rho} = \sum_j \rho_j | \mu_j \rangle\langle \mu_j |.
\end{equation}

For the system sizes that we consider, the $\mathcal{O}(\Omega)$ minimization problem can be solved quickly by standard numerical minimization packages. This is the approach we use in the main text of this article. For larger problems, Smolin et al. provide a simple algorithm after reducing the complexity of the problem further by noting that the solution to \cref{eqn:ml-algo-2} essentially involves finding a `pivot' $j^\prime$ in the (ordered) $\mu_j$ wherein $\rho_j = 0$ for $j < j^\prime$ and $\rho_j = \mu_j + c$ where $c$ is a constant for $j \geq j^\prime$. The use of this algorithm is unnecessary for the situations we consider in this article, and we refer the interested reader to \cite{smolin2012efficient} for more details.

\end{document}